\documentstyle[aps,psfig]{revtex}
\def\nl{\noindent}

\def\nn{\nonumber}

\def\Journal#1#2#3#4{{#1} {\bf #2} (#3) #4}
\def\NCA{\em Nuovo Cimento}

\def\NPB{{\em Nucl. Phys. B}}

\def\PRL{\em Phys. Rev. Lett.}

\title{ Crossing Symmetry  Violation of  Unitarized \\ Pion-Pion Amplitude in
the Resonance Region}
\author{Isabela P. Cavalcante\thanks{E-mail: ipca@uerj.br} \ and
J. S\'a Borges\thanks{E-mail: saborges@uerj.br}} 
\address{ Universidade do Estado do Rio de Janeiro\\ 
Rua S\~ao Francisco Xavier, 524, Maracan\~a, Rio de Janeiro, Brazil}
\begin{document}

\maketitle

\begin{abstract}
 
Pion-pion scattering amplitude obtained from one-loop Chiral
Perturbation Theory (ChPT)  is crossing symmetric, however the
corresponding partial wave  amplitudes do not respect exact
unitarity relation. There are different approaches to get 
unitarized partial wave amplitudes  from ChPT. Here we consider  the inverse
amplitude method (IAM) that is often used  to fit pion-pion
phase shifts  to experimental data, by adjusting free parameters.   
We  measure the amount of crossing
symmetry violation  (CSV) in this case and 
we show that  crossing symmetry is badly violated by the IAM
unitarized ChPT amplitude in the resonance region. Important CSV also
occurs 
when  all  free parameters are set equal to zero. 
\end{abstract}

\vspace{.5cm}

\nl

\section{Introduction}

Even though Quantum Chromodynamics (QCD) has achieved a great success
in describing strong interactions, low energy hadron physics must
still be modeled phenomenologically.  
A great theoretical improvement was made by means of the method of
ChPT \cite{Leu}, which is an effective theory
derived from the basis of QCD.  The method consists of writing down chiral
Lagrangians for the physical processes and  uses the conventional
technique of the field theory for the calculations. 

Here we focus on the simplest interaction to test ChPT ideas,
namely the   pion-pion
scattering. For this process, 
the   ChPT leading  contribution  (tree graphs) is of second order in the
momenta $p$ of the external pions and coincides with   Weinberg 
result from  current algebra \cite{Wei}.
 The corrections come from  loop diagrams
whose  vertices are of order $p^2$ and include a free-parameter polynomial 
part related to 
tree diagrams of order $p^4$; these
parameters have to be  obtained phenomenologically.
At each order of calculation, the method yields a total
amplitude that respects exact crossing symmetry, however, the corresponding 
partial waves satisfy only approximate elastic unitarity relation.

The unitarity violation is more severe at higher energies,  
so that it is not possible to
reproduce resonant states, which are one of the most relevant features
of the strong interacting regime. This is not a new issue in
literature and
many different methods have been
proposed to improve this behavior.
Here we  consider 
IAM \cite{dob}, that 
 implements exact elastic unitarity
for $S-$ and $P-$partial waves. It
allows 
one to access the resonance region for pion-pion scattering by
fixing free parameters, 
but violates crossing symmetry.  

Crossing symmetry interconnects the various isospin channels.
In other words, if one fits IAM isospin $I=1$ amplitude 
 to the  experimental data up the the $\rho$ resonance
 region, for instance, due to crossing symmetry we do not have much
freedom for $I=0$ fitting.
 Therefore, if crossing  symmetry is not fulfilled, a very 
constrained fit of $S-$wave would give  meaningless result.  
It is thus important to investigate at what extent  CSV 
occurs. 

A possible test for crossing symmetry is the use of the so called Roskies
relations \cite{Ro}, which involve  
weighted integrals of  $S-$ and $P-$waves in
 the nonphysical region ($4 m_\pi^2 > s >
 0 $) and follow from
 first principles of analyticity, crossing and unitarity.
Another test is given by the so called Martin
inequalities \cite{Mar}. They are constraints on   
$\pi^0 \pi^0$ scattering amplitude also in the 
nonphysical region and follow from analyticity and positivity of
the imaginary part of the amplitude.
Both methods have been used in literature and
 show that Roskies relations violations are below 1.3\% and all Martin
inequalities are satisfied by low energy ChPT IAM pion-pion
 amplitude at one loop approximation level \cite{tobe}.
 In fact,
 as unitarity corrections are small, we do 
 expect small  CSV at very low energies. 

At this point, we would like to emphasize that  IAM
 aims to explore  the resonance region, where large unitarity
 corrections are needed and, accordingly, sizable
CSV is expected.  Therefore, in the present paper, we  extend
the test to that  region. We do that by evaluating the effect of the
 IAM modified $S-$
 and $P-$waves on the 
exact crossing symmetry relation for the total amplitude.

Our work is presented as follows.
In Sect. \ref{sec:iam} we write the ChPT amplitude for pion-pion
scattering and we construct IAM partial waves. 
We introduce  a correction to get  rid of sub-threshold poles 
by slightly shifting the original Adler zeros of the
 leading amplitudes.
In Sect. \ref{sec:method} we present the method used to quantify
the CSV of IAM amplitudes. It consists in  
comparing two expressions of the total amplitude,
obtained from different combinations of isospin partial
waves. In this section we define a violation function which is
calculated and plotted for some values of energy and scattering angle.
In Sect. \ref{sec:con} we 
present our conclusions.

%%%%%%%%%%%%%%%%%%%%% SECTION II %%%%%%%%%%%%%%%%%%%%%%%%%%%%%%%%%%%%%%%%%
\section {Chiral perturbation theory  and the IAM}
\label{sec:iam}

We consider $\pi^a \ \pi^b \rightarrow \ \pi^c \ \pi^d$ scattering
amplitude 
 $$ \langle \pi^c \pi^d \vert \ \ T \ \  \vert \pi^a \pi^b
\rangle = A(s,t,u) \delta^{a b} \delta^{c d} + B(s,t,u) 
\delta^{a c} \delta^{b d} +  C(s,t,u) \delta^{a d} \delta^{b c}.
$$
The total isospin defined amplitudes  $T_I$\, for $I\, =\,  0, 1\,\,
{\hbox{and}}\, \,  2$ are
\begin{eqnarray}
T_0 (s,t) &=& 3 A(s,t,u)\, +\, B(s,t,u)\, + \, C(s,t,u) \, , \nonumber\\
T_1 (s,t) &=&  B(s,t,u)\, - \, C(s,t,u) \, , \label{isos}\\
T_2 (s,t) &=&  B(s,t,u)\, + \, C(s,t,u), \nonumber
\end{eqnarray}
Crossing symmetry  implies that
there is just one amplitude 
describing the three total isospin channels of the process, so that 
\begin{eqnarray}
B(t,s,u)= A(s,t,u) = C(u,t,s)\label{cros}.
\end{eqnarray}
The  amplitudes  $T_I$ are
expanded in partial waves, as
\begin{eqnarray}
T_I(s,t)\, = \, \sum_{\ell} (2 \ell + 1 ) \, t_{\ell\, I}(s) \, P_\ell \, (\cos
\theta), \label{ti}
\end{eqnarray}
\nl where $P_\ell$ are the Legendre polynomials, $2 t = (s - 4
m_\pi^2)(\cos \theta - 1)$, and $u = 4 m_\pi^2 - s - t$. 
Elastic unitarity implies that,  for\  $    16 m_\pi^2 \ge s \ge 4
m_\pi^2$,\  there is a constraint given by  
$$
{\hbox{Im}} \, t_{\ell \, I} (s) = \rho (s) |  t_{\ell \, I} (s) |^2,
$$
\nl which can be solved yielding 
\begin{equation}
 t_{\ell \, I} (s) = \frac 1 {\rho
(s)} e^{i\, \delta_{\ell \, I}(s)}\, \sin \delta_{\ell \, I}(s),
\label{delta}
\end{equation}
where $ \delta_{\ell \, I}(s)$ are the real phase shifts and
$$\rho(s) = \frac 1 { 16 \pi} \sqrt{\frac{s - 4 m_\pi^2}{s}}$$ 
is the phase space
factor for pion-pion scattering. 

Using ChPT
at the one-loop level and considering only the most relevant low
energy constants,  the amplitude is  given by
\begin{eqnarray*}   
f_\pi^4 A(s,t,u)& = & f_\pi^2 ( s - m_\pi^2 ) +  
\frac 1 2 \,\left (s - \, m_\pi^2 \right )^2 \bar J(s) \\
&+& \left[ \frac 1 {12}\, \left (3\,\left (t-2\,{m_\pi}^{2}
\right )^{2}+\left (s-u\right )\left (t-4\,{m_\pi}^{2}\right )\right
)\bar J(t) + (t \leftrightarrow u) \right]  \\ &+& 
\lambda_{{1}}\left (s-2\,{m_\pi}^{
2}\right )^{2}+\lambda_{{2}}\left[ \left(t - 2\,{m_\pi}^{2}\right)^{2}
 + \left(u - 2\,{m_\pi}^{2}\right)^{2} 
\right].
\end{eqnarray*}

The resulting  
ChPT amplitudes for  $S-$wave ($I = 0\ {\hbox{and}}\ 2 $)   and  $P-$wave ($I =1 $)
 can be written as
\begin{equation}
 t_{\ell \,I} (s) = t_{\ell \,I}^{ca}(s) + t_{\ell \,I}^{ca\ ^2}(s)\,  \bar J(s) +
t_{\ell \,I}^{left}(s) +p_{\ell \,I}(s),
\label{chpt}
\end{equation}
where $t_{\ell \,I}^{ca}$
are the (real) Weinberg amplitudes, namely,
\begin{eqnarray*} 
f_\pi^2 \, t_{0 0}^{ca}(s) = 2 s - m_\pi^2, \quad f_\pi^2\, t_{1 1}^{ca}(s) =
\frac 1 3\, (
s - 4 m_\pi^2),\quad  f_\pi^2\, t_{ 0 2}^{ca}(s) =  2 \, m_\pi^2
- s,
\end{eqnarray*}
\  $t_{\ell \,I}^{left}$ are  the parts that bear
the left-hand cuts, namely,
\begin{eqnarray*}
f_\pi^4 \pi^2  \, t_{0 0}^{left}(s)&=&\frac 1 {12} \frac {m_\pi^4}{s-4
m_\pi^2} ( 6 s - 25 m_\pi^2)\, L(s)^2 - \frac 1 {72 \rho(s)} ( 7 s^2 -
40 m_\pi^2 s + 75 m_\pi^4) L(s)  \\ 
&+& \frac 1 {864} ( 95 s^2 - 658 m_\pi^2
s + 1454 m_\pi^4),\\ \\
f_\pi^4 \pi^2  \, t_{0 2}^{left}(s) &=& - \frac {1} {12} \frac {m_\pi^4}{ s  - 4
m_\pi^2} ( 3 s + m_\pi^2)\, L(s)^2 - \frac 1 {144 \rho(s)} ( 11 s^2 -
32 m_\pi^2 s + 6 m_\pi^4) L(s)\\ &+& \frac 1 {1728} ( 157 s^2 - 494 m_\pi^2
s + 580 m_\pi^4), \\
\lefteqn{f_\pi^4 \pi^2 ( s - 4 m_\pi^2)\,  t_{1 1}^{left}(s) \, = \, \frac 1 {12} \frac {m_\pi^4}{s-4
m_\pi^2} ( 3 s^2  - 13 m_\pi^2 s - 6 m_\pi^4)\, L(s)^2 } \hspace*{2cm} \\ 
&+& \frac 1 {144 \rho(s)} (  s^3 -
16 m_\pi^2 s^2 + 72 m_\pi^4 s -36 m_\pi^6) L(s) \\ &-& \frac 1 {864} ( 7
s^3 - 71 m_\pi^2  s^2 + 427  m_\pi^4 s - 840 m_\pi^6),\quad
{\hbox{with}}
\end{eqnarray*}
$$\bar J (s) = \frac 1 {8 \pi^2} -\frac 2 {\pi}\,  \rho (s)\, \,  L(s)
+ I\, 
\rho (s) \, , \quad  \quad
L(s)\,  =\,  \ln \, \frac{\sqrt{s-4 m_\pi^2} + \sqrt{s}}{2 m_\pi} \, ,$$ 
 and $ p_{\ell \, I} (s)$\ are two free parameter polynomials,
given by
\begin{eqnarray*}
f_\pi^4 \, p_{0 0} (s) &=& 
\frac 1 3 \, 
\left (11\,{s}^{2}-40\,s{m_\pi}^{2}+44\,{m_\pi}^{4}\right
)\lambda_{{1}}+\frac 1 3 \left (14\,{s}^{2}-40\,s{m_\pi}^{2}+56\,{m_\pi}^{4}\right
)\lambda_{{2}}, \\
f_\pi^4 \, p_{1 1} (s) &=& \frac 1 3 s \left(s - 4 m_\pi^2 \right) \left( \lambda_2
- \lambda_1\right) \, ,\\
f_\pi^4 \, p_{0 2} (s) &=& 
\frac 2 3 \,
\left ({s}^{2}-2\,s{m_\pi}^{2}+4\,{m_\pi}^{4}\right )\lambda_{{1}}+
\frac 2 3 \, \left
(4\,{s}^{2}-14\,s{m_\pi}^{2}+16\,{m_\pi}^{4}\right )\lambda_{{2}} \, .
\end{eqnarray*}

If one wants to describe a resonant amplitude, one may wish to use
Pad\'e approximants,  as e.g. advocated in \cite{dob}. It amounts to
writing the inverse of the  partial wave. Thus,  instead of
the exact ChPT result $t_{\ell \, I}$, we use a modified 
amplitude 
\begin{equation}
\tilde t_{\ell \, I} (s) = \frac {t_{\ell \,I}^{ca}(s)}{ 1 -  \left( t_{\ell \,I}^{ca\ ^2}(s)\,  \bar J(s) +
t_{\ell \,I}^{left} (s) + p_{\ell \,I}(s)\right)/t_{\ell
\,I}^{ca}(s)}.
\label{ttil}
\end{equation}

\begin{figure}[!]%[h]
\centerline{\psfig{figure=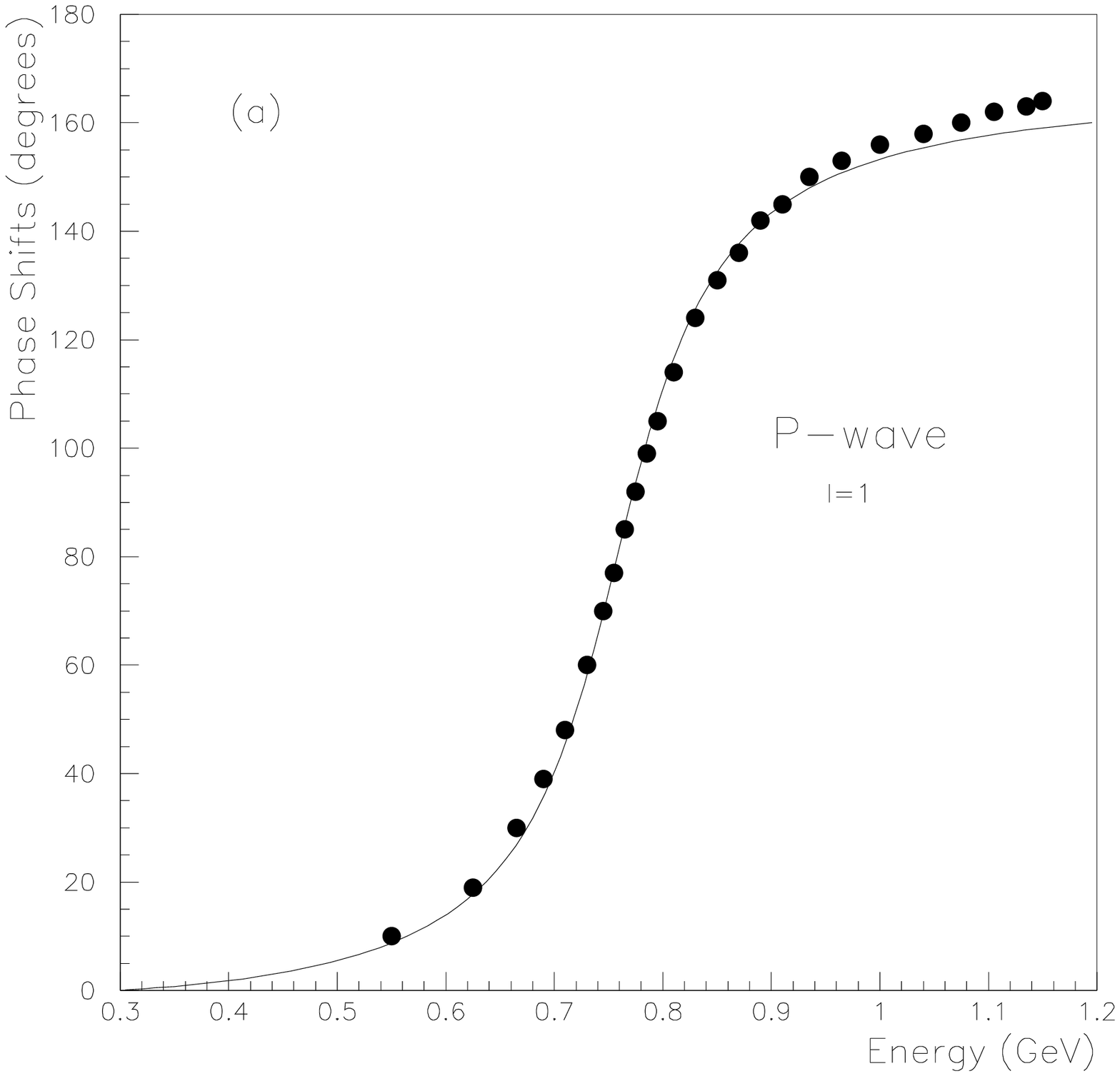,height=8.0cm}\hspace{0.1cm}
\psfig{figure=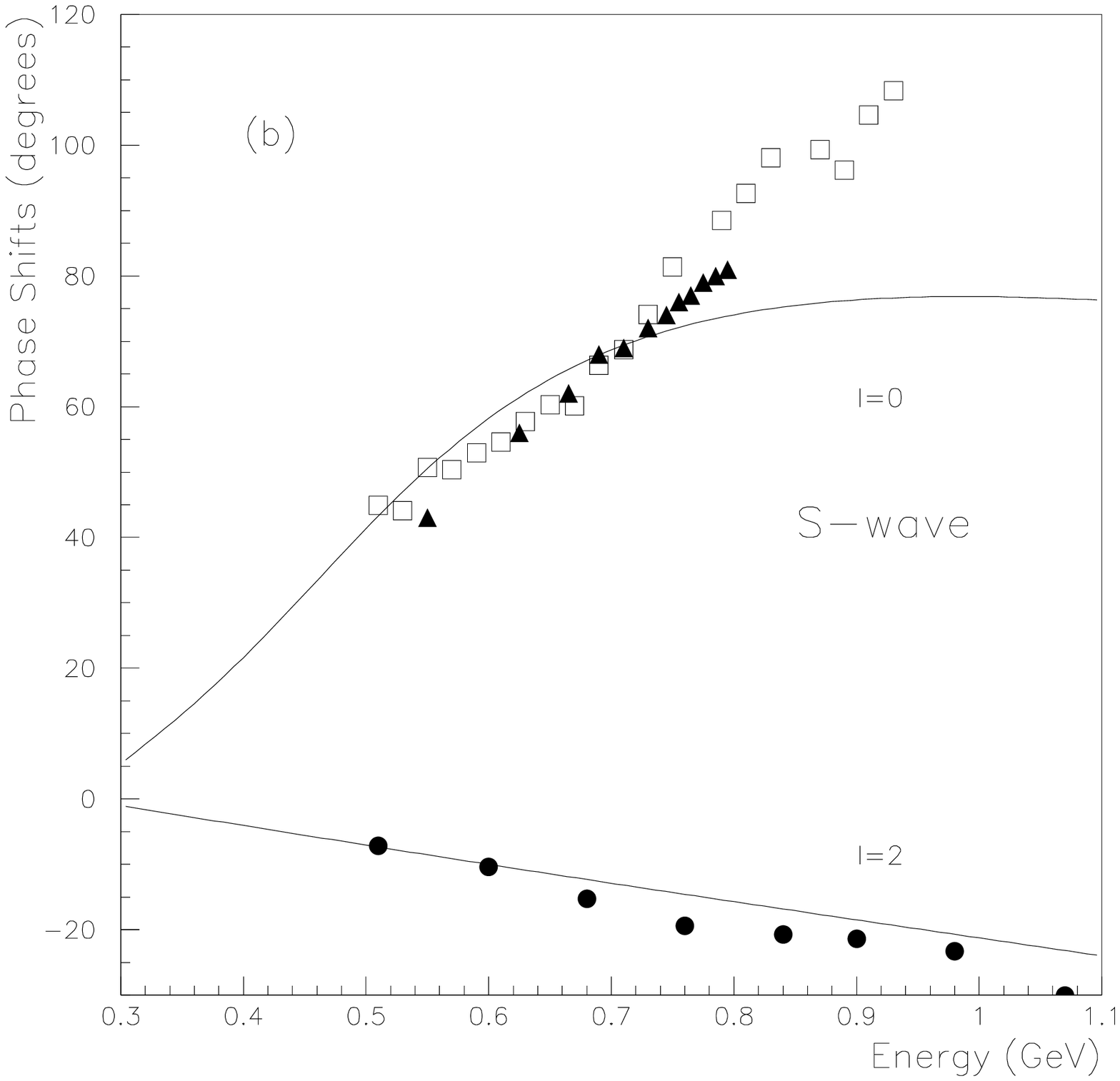,height=8.0cm}}
\caption{Results from fits of IAM amplitudes 
to (a) $P-$ and  (b) $S-$wave phase shifts, in
degrees, as functions of cms energy, in GeV. Experimental data for
$P-$wave are from
Ref. [9];  for $S-$wave, from Refs. [9-11].}
\label{iam_fits}
\end{figure}

We employ here a strategy widely used in literature \cite{dob},
namely
we choose the parameters $\lambda_1$\ and $\lambda_2$
in order to fit S- and P-waves  above to the experimental phase shifts,
by using the definition (\ref{delta}).
We show in Fig. \ref{iam_fits} the resulting phase shifts
corresponding to the parameters $\lambda_1 = -0.00345$ and $\lambda_2
= 0.01125$\cite{isa}.

As mentioned in the introduction, there was a problem concerning
S-waves, namely that they were singular  
 at some sub-threshold value for $s$, where  the correction becomes equal to
$t^{ca}$. 
Singularities occur in S-wave sub-threshold amplitudes 
at $s_0\simeq 0.64 \, m_\pi^2$, for $I=0$, and at
$s_2 \simeq 1.95 \, m_\pi^2$,
for $I=2$. Those values are  close to the ones where
$t_0^{ca}$ and $t_2^{ca}$ actually vanish. 
On the other hand, if one takes $\lambda_1 = \lambda_2 = 0$ the singularities
move to  $s_0\simeq 0.51 \, m_\pi^2$, for $I=0$, and to
$s_2 \simeq 1.99 \, m_\pi^2$,
for $I=2$.
In order to get rid of those singularities,  we performed an extra
correction, thus obtaining a new partial wave amplitude, denoted by \,
$\tilde t_I^{(n)}$,
\begin{equation}
\tilde t_I^{(n)} (s) = \frac {\alpha_I (s-s_I)/f_\pi^2}{ 1 -  \left( t_I^{ca\ ^2}(s)\,  \bar J(s) +
t_I^{left} (s) + p_I(s)\right)/t_I^{ca}(s)}, \quad I=0\ {\hbox{and}}\
2 ,
\label{tn}
\end{equation}
where $\alpha_0 = 2$ and  $\alpha_2 = -1$.

%%%%%%%%%%%%%%%%%%%%% SECTION IV %%%%%%%%%%%%%%%%%%%%%%%%%%%%%%%%%%%%%%%%%
\section{Crossing symmetry violation}
\label{sec:method}

We present here a method to 
quantify CSV of the IAM pion pion scattering
amplitude in the resonance region. 
Since IAM only modifies $S-$ and $P-$waves, we start by separating in
Eq. \ref{ti} the $\ell = 0$ and $\ell =1$ contributions:
\begin{eqnarray}
T_0(s,t) &=& t_{00}(s) + \sum_{\ell =2}^\infty (2 \ell +1) t_{\ell 0}(s)
P_\ell(\cos \theta),\\
T_2(s,t) &=& t_{02}(s) + \sum_{\ell =2}^\infty (2 \ell +1) t_{\ell 2}(s)
P_\ell(\cos \theta),\\
T_1(s,t) &=& 3 \, t_{11}(s)\ \cos \theta + \sum_{\ell =3}^\infty (2 \ell +1) t_{\ell 1}(s)
P_\ell(\cos \theta).
\end{eqnarray}
\label{Tet}
In general, the
total amplitudes $A(s,t)$ and $B(s,t)$ can be reconstructed from
the isospin defined amplitudes above, 
considering 
Eqs. (\ref{isos}), which gives
\begin{eqnarray}
 A(s,t) & = &  \frac 1 3 \left[ T_0 (s,t) - T_2 (s,t)\right],\nonumber \\
B(s,t) & = &   \frac 1 2 \left[ T_1 (s,t) + T_2 (s,t)\right].
\label{asat}
\end{eqnarray}
Thus we have
\begin{eqnarray}
\tilde A(s,t) &=& 
\frac 1 3 \left( \tilde t_{00} (s) - \tilde t_{02} (s) \right) +  
\left[ A(s,t) -  \frac 1 3 \left( t_{00} (s) - t_{02} (s) \right)\right], \nn \\
\tilde B(s,t) &=& 
\frac 1 2 \left( 3 \,\tilde t_{11} (s)\, \cos \theta  + \tilde t_{02} (s)
\right) +\left[  B(s,t) -
\frac 1 2 \left(  3 \, t_{11} (s)\, \cos \theta  +  t_{02} (s)
\right) \right], 
%\hspace{3cm} 
\nn
\end{eqnarray}
where the tilde indicates that these amplitudes are built from the IAM
modified partial waves.

According to Eq. \ref{cros}, if crossing symmetry were respected, one
should have $\tilde B(s,t) =  \tilde A(t,s)$. Thus the difference
between these two quantities is a measure of the symmetry
violation. 
We remark that $B(s,t)$ is a combination of $P-$ and $S-$waves while
$A(t,s)$  is reconstructed from  $S-$wave amplitudes only.
In order to put some scale in this measure, we divide this
difference by their sum, so that the violation is defined as
\begin{equation}
\Delta (s,\cos \theta) = 100 \% \times \frac {\vert \tilde B(s,t) - \tilde A(t,s)\vert}{
|\tilde B(s,t)|+| \tilde A(t,s)|}.
\end{equation}
This formula consists in a
simple way to translate CSV into numbers. 
A similar formula was used by Boglione and Pennington \cite{bo97} and by
Hannah \cite{tobe} to quantify the violation of Roskies relations in
pion-pion scattering amplitude, mentioned in the introduction. We
modified the denominator in order to constraint the amount of violation
to 100\%.

We have shown  that data
fitting requires  the  partial wave polynomial parts to be adjusted.
However, 
as IAM represents a structural modification of
the amplitudes,
it should violate crossing independently of its 
polynomial part.
Therefore, let us first consider the case  $\lambda_1
= \lambda_2 = 0$.

\begin{figure}[!]%[b]
\centerline{\psfig{figure=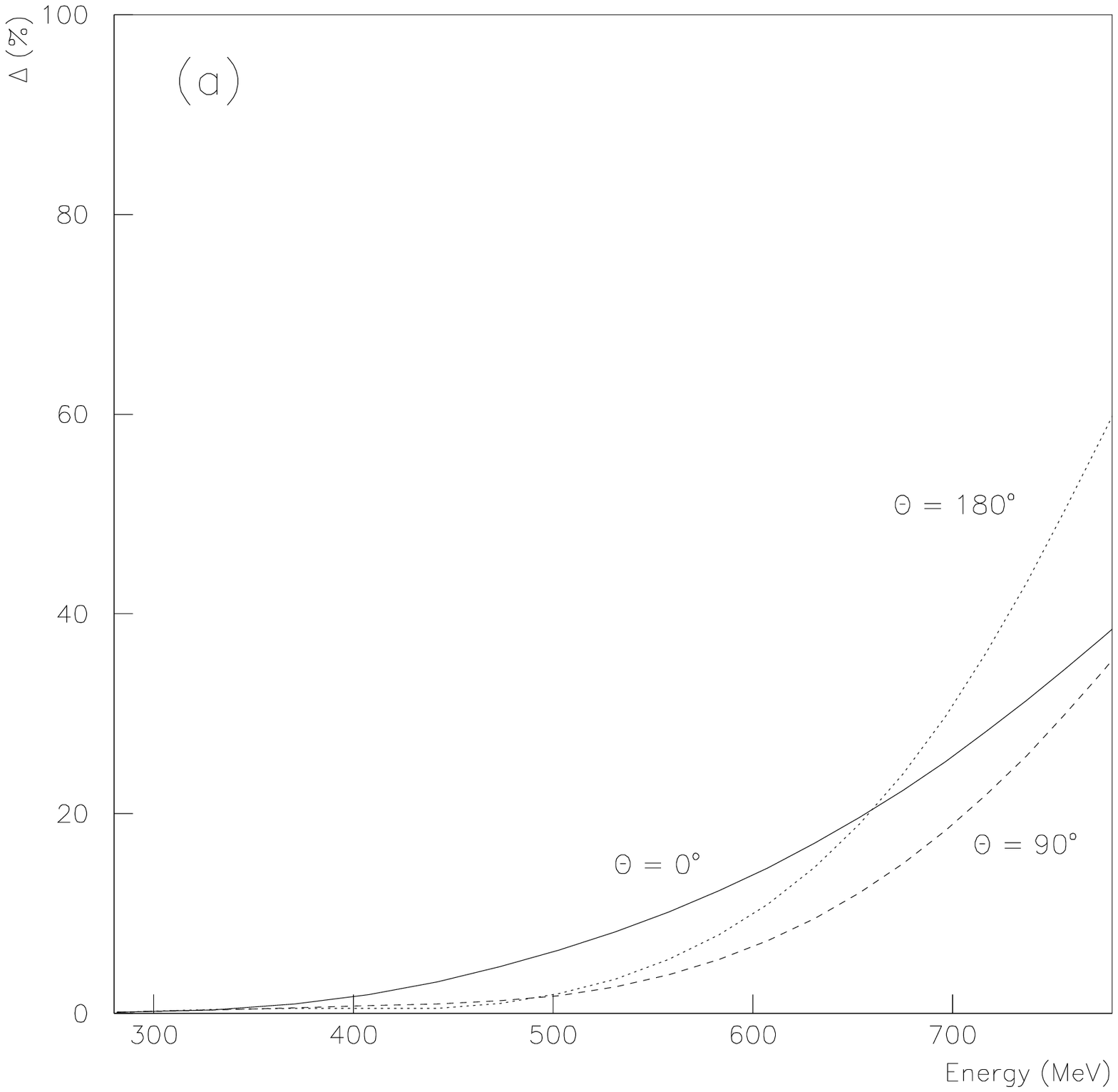,height=8.0cm}%\hspace{0.1cm}
\psfig{figure=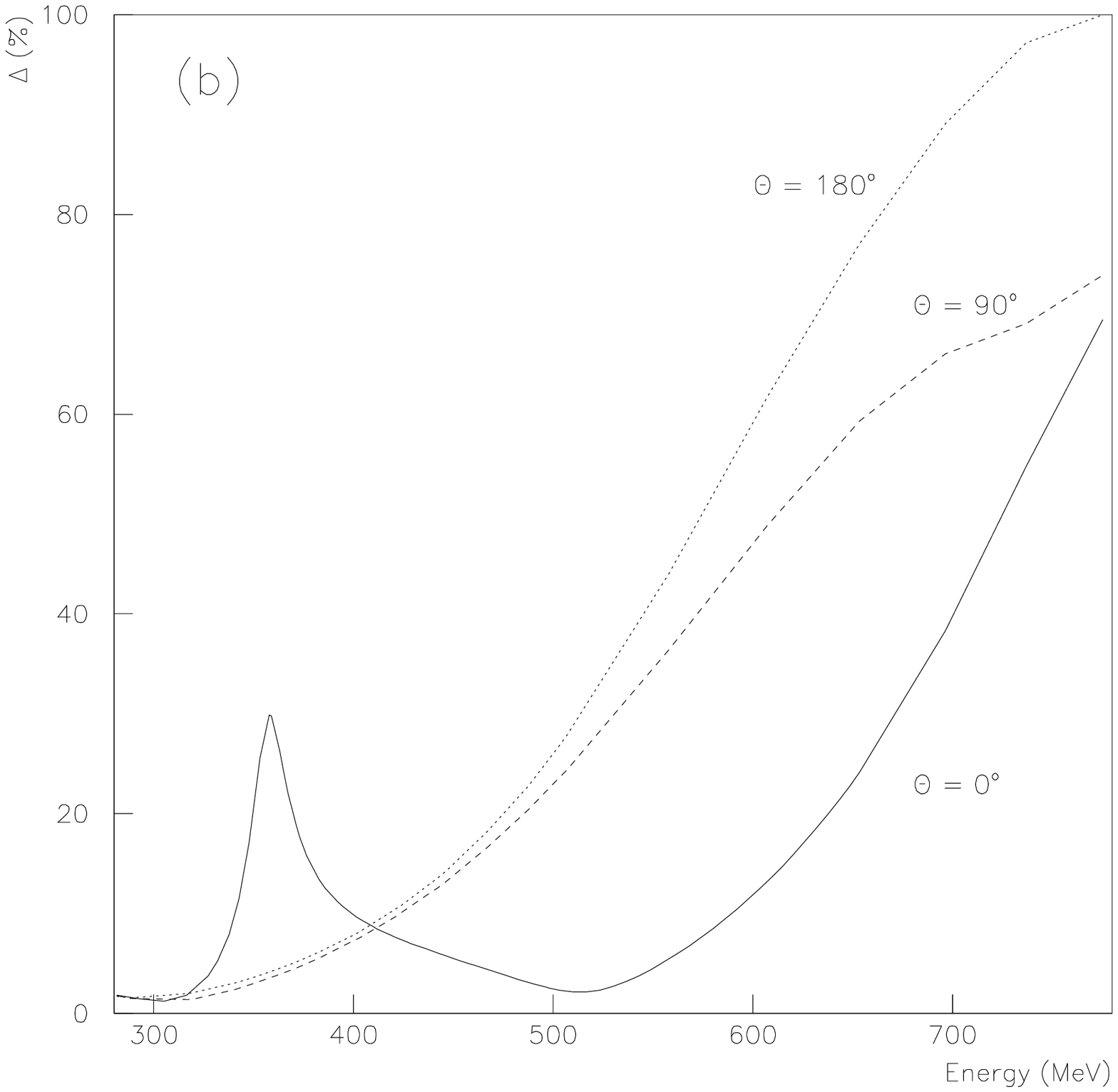,height=8.0cm}}
\caption{Crossing symmetry violation as a function of cms energy, in
MeV, for
$\theta = 0^0$  (solid),  $ 90^{0}$ (dashed) and $ 180^{0}$
(dotted) for (a) $\lambda_1=\lambda_2=0$  and for 
(b)  $\lambda_1 = -0.00345$ and $\lambda_2 = 0.01125$. }
\label{viola-fig}
\end{figure}

In Fig. \ref{viola-fig} we have plotted $\Delta$ for some values of
scattering angle as functions of cms energy, for the cases with
vanishing parameters and with adjusted values. 
 For the case when
the parameters are kept equal to zero, the curves show the same trend,
while, when the adjusted
parameters are employed, CSV for the forward
scattering presents a peculiar  behavior. Concerning the strength of
violation, one notices that the use of adjusted parameters increases $\Delta$.

%%%%%%%%%%%%%%%%%%%%% SECTION V  %%%%%%%%%%%%%%%%%%%%%%%%%%%%%%%%%%%%%%%%%
 \section{Conclusions}
\label{sec:con}

The  $\cal O$(p$^4$) ChPT pion-pion amplitude is crossing symmetric
but does  not
respect exact elastic unitarity. There are several attempts to
extrapolate the domain of validity of 
ChPT and to access the resonance region for meson-meson
scattering. IAM is one  of these methods, which  
fixes some free parameters
 in order to fit Pad\'e approximants of  ChPT amplitude  to the experimental
data. The parameters are fixed by means of a simultaneous fit of
$S-$ and $P-$waves. One knows that the interdependence of the fits  
relies on the interconnection among different isospin amplitudes,
which is due to  crossing symmetry.
However,
IAM leads to unitary amplitudes, but they violate crossing symmetry.  
As a consequence, the parameter fitted from $P-$wave may be
meaningless for the $S-$wave fit, since the constraint among partial waves
is somehow lost. In this sense, the amount
of CSV is an indication of the lost of
reliability of the fits performed. 

CSV of IAM has been quantified before by means of the Roskies relations
and Martin inequalities. Both methods found very small violations at
sub-threshold energies. 
In this paper we presented a method to quantify the CSV
 that IAM, extended to the resonance region, implies. 

We measured the violation for two cases. 
As IAM produces a
structural modification of ChPT results, any choice of parameters will
yield CSV. Thus 
we first considered $\lambda_1 =
\lambda_2 = 0$ and as a result we obtained very small CSV at
threshold, as expected, while values up to roughly 50\% develop around the
$\rho$ mass region, for any scattering angle, as shown in
Fig. \ref{viola-fig}a. This
can be considered as a measurement of the intrinsic CSV of IAM.
On the other hand, using
the parameters fixed in order to fit the amplitudes to experimental
data, the violation is still small near threshold, but  gets much
larger at higher energies, as shown in the Fig. \ref{viola-fig}b.
This result can be taken as a measure of the price one pays for
imposing elastic unitarity on ChPT amplitudes far from threshold.

Our results show that it is not possible for ChPT to
exactly fulfill both 
crossing symmetry and unitarity 
 requirements. We recall that  elastic
unitarity constrains partial waves in a small energy range (in this
case, up to 560 MeV) in contrast with crossing symmetry of the total
amplitude.
An alternative method to keep exact crossing symmetry is to fit 
pure ChPT partial wave amplitudes in Eq. \ref{chpt} 
to experimental
data \cite{bor}. In this case 
very large unitarity violations in the resonance region occur. 
In other words, by introducing elastic unitarity, a
lot of crossing symmetry is lost, as well as keeping the latter\cite{bor} costs
a large amount of the former.

\section*{Acknowledgment}

The work by I.P.C. was supported by CNPq (Brazilian agency), grant
number 150194/99-4.

\end{document}